# Switchable collective pinning of flux quanta using "magnetic vortex" arrays

J.E. Villegas[1,2], K.D. Smith[2], Lei Huang[3], Yimei Zhu[3], R. Morales[2] and Ivan K. Schuller[2]

[1]*Unité Mixte de Physique CNRS/Thales, Route Départementale 128, 91767 Palaiseau, France and Université Paris Sud 11, 91405 Orsay, France*

[2]*Physics Department, University of California – San Diego, 9500 Gilman Drive, La Jolla CA 92037-0319*

[3]*Department of Condensed Matter Physics and Materials Science, Brookhaven National Laboratory, Upton, NY 11973*

We constructed a superconducting/ferromagnetic hybrid system in which the *ordering* of the pinning potential landscape for flux quanta can be manipulated. Flux pinning is induced by an array of magnetic nanodots in the "magnetic vortex" state, and controlled by the magnetic history. This allows switching *on* and *off* the collective pinning of the flux-lattice. In addition, we observed field-induced superconductivity that originates from the annihilation of flux quanta induced by the stray fields from the "magnetic vortices".

# I. INTRODUCTION

The interaction between elastic lattices and fixed pinning potentials is a problem common to a variety of physical systems, e.g. repulsive colloidal particles [1] and Bose-Einstein condensates [2] in optical lattices, charge density waves in solids [3] or flux quanta (Abrikosov vortices) in type-II superconductors [4-12]. The phase diagram, ordering and dynamics of these systems are strongly influenced by that interaction, and ultimately by the geometry and degree of order of the pinning substrate [10,13]. This is dramatically illustrated by flux-lattice dynamics with artificial pinning potentials, where commensurability with *periodic* [4-11] and *quasiperiodic* [12] potentials induces *collective* or *local* pinning [14] and controls lattice correlation lengths. In this context, the realization of a system where the ordering of the pinning potential can be switched by an external parameter is especially interesting.

Ordered arrays of magnetic nanoparticles (dots or other geometric nanostructures) have been widely used to create pinning potentials for the flux-lattice in superconducting thin films [5-7,9-11]. In addition to the "structural pinning" (observed also in nonmagnetic structures like arrays of antidots [8]), the magnetic character of the nanoparticles generates several pinning mechanisms [9]. These include proximity effect [15], magnetic reversal losses [16], and magnetostatic interactions between flux quanta and the stray magnetic fields from the nanoparticles [17]. If the latter is the governing mechanism, the pinning potential strongly depends on the magnetic state of the nanoparticles. This gives rise to asymmetric (field polarity dependent) [6,7] flux pinning and to pinning potentials of tunable strength [11]. These effects have been observed in arrays where the individual nanoparticles present a virtually identical *magnetic multidomain state* and the same remanent magnetization $\vec{M}$. Thus, the interaction between flux quanta and every single magnetic particle in the array is virtually identical, *and changes in $\vec{M}$ do not affect the ordering of the pinning potential*, which is fixed by the array geometry.

In this article we report on a system in which the collective pinning of flux quanta can be switched *on* and *off*, as opposed to the systems mentioned above in which the ordering of the potential, and hence the occurrence of collective pinning, is fixed. This is achieved by manipulating the magnetic order of a periodic array of dipoles arranged on top of a superconducting film, as depicted schematically in Fig. 1. If the array is magnetized (all the dipoles point in the same direction), a periodic pinning potential is obtained, which causes collective pinning and induces square symmetry order in the flux-lattice. If the array is demagnetized (balanced distribution of dipoles pointing in opposite directions), a disordered pinning potential is obtained, and no commensurability develops between the flux-lattice and the array. In addition to this effect, we have observed field-induced superconductivity [18,19], which originates here from the annihilation of dipole-induced flux quanta [20]. The manipulable array of magnetic dipoles was realized using ordered arrays of magnetic dots in the so-called "magnetic vortex" state [21-26]. This system is similar to that used earlier to induce bistable superconductivity in thin films [26], except in the arrays studied here dot sizes and inter-dot distances are larger than the superconducting coherence length, giving rise to different Physics. More elaborate extensions of this experimental realization could be used to create magnetic pinning potentials of tunable geometry, asymmetry, etc.

**II. SAMPLE CHARACTERIZATION**

Samples consist of square arrays of Co dots (Fig. 2 (a)), either directly on top of 30 nm thick $Si_3N_4$ membranes for Lorentz microscopy experiments, or on top of Al/AlO$_x$ bilayer thin films for transport experiments. For the latter, after Al evaporation onto sapphire substrates, the films were exposed to air in order to obtain a ~3 nm thick native AlO$_x$ capping layer. Dot arrays were defined on a 50 μm × 50 μm area using e-beam lithography, sputtering, and lift-off techniques [27]. Dots consist of two layers, Co (40 nm thick) and Au (2 nm thick, to prevent Co oxidation). Several square arrays with interdot distance $a$ =0.6-1 μm and dot diameters $\varnothing$ =430-490 nm were fabricated. A 40 μm × 40 μm (long × wide) four-probe standard bridge for transport measurements was optically lithographed. For $10-25$ nm thick Al/AlO$_x$ films, superconducting critical temperatures were respectively $T_c = 1.95-1.65$ K, coherence lengths $\xi(0) \approx 40-50$ nm (estimated from upper critical fields $H_{c2}$), and penetration depths $\lambda(0) \approx 350-220$ nm (estimated from $T_c$ and the residual resistivity [28]). Therefore $\kappa(0) = \lambda(0)/\xi(0) \approx 8.5-4.5$, i.e. the Al films studied here are type-II superconductors.

The aspect ratio $\varnothing/h$ of the Co dots (with $h = 40$ nm the dot height) was chosen [24] so that their magnetic ground-state is a "magnetic vortex" [29]. In this, the magnetization curls *in-plane* clockwise or counter-clockwise (vortex *chirality*) around a core, where it points up or down *out-of-plane* (vortex *polarity*). Fig 2 (a) shows a Lorentz image of a demagnetized array, in which the "magnetic vortex" cores appear as black (white) spots in the center of the dots for clockwise (counter-clockwise) chirality [30,31]. Further evidence of this "vortex-state" arises from *in-plane* hysteresis loops (Fig. 2 (b)). These present a pronounced "pinching" in the middle of the loop, characteristic [21,25] of magnetic reversal via nucleation, displacement, and annihilation of "magnetic vortices". A cartoon of this reversal is shown in Fig. 2 (b): from negative to positive saturation (coded red (darker gray) to blue (lighter gray)), those three consecutive events are depicted. Because of the flux-closure distribution of the *in-plane* magnetization, the stray magnetic field from these nanodots is essentially produced by the *out-of-plane* magnetic moment of the vortex core, which resembles a magnetic dipole (inset in Fig. 2 (b)). As experimentally shown earlier for similar arrays of "magnetic vortices" [22,23], all of them have the same *polarity* in the remanent state (i.e. vortex cores throughout the array have parallel magnetization) after application and removal of a sufficiently intense *out-of-plane* field. On the other hand, after an *out-of-plane* demagnetizing cycle, the distribution of "vortex polarities" is balanced (there is an equal number of cores with magnetization pointing up/down) [22,23]. Contrary to other systems in which demagnetizing the array causes the demagnetization of each individual dot [19,32], here each dot keeps a permanent magnetic moment (the "vortex core"). Thus, these arrays of "magnetic vortices" constitute a realization of the scenario in Fig. 1. Moreover, since the insulating $AlO_x$ layer strongly reduces the proximity effect [15] between Co dots and the superconducting Al film, the magnetostatic interaction between "magnetic vortex" cores and flux quanta is the governing pinning mechanism.

### III. RESULTS AND DISCUSSION

Fig. (3) shows the mixed-state magnetoresistance of one of the samples (Al thickness 10 nm, dot array with $a = 0.6$ μm and $\varnothing = 490$ nm), with the external magnetic field $H$ applied *out-of-plane*, at $T = 0.87 T_c$ and for several injected current levels (see legend). The rest of the samples show a similar behavior. *This behavior differs depending on the magnetic state of the array*.

Fig. 3 (a) corresponds to the case where a field $H = 20$ kOe was applied perpendicular to the film plane prior to $R(H)$ measurements. Therefore, in the remanent state all the vortex cores have parallel magnetization throughout the array, leading to a situation as in Fig. 1 (a). Note that this state remains unaltered during $R(H)$ measurements since $H < 400$ Oe, well below the field strength needed to reverse the core magnetization (typically several kOe [22,23]). Three main features are remarkable. First, the absolute minimum of $R(H)$ is not at $H = 0$, but shifted to a field $H_S \approx 25$ Oe. This corresponds to a magnetic flux $H_S a^2 \approx 0.5 \phi_0$ per unit cell of the square array, with $\phi_0 = 2.07 \cdot 10^{-7}$ Mw the flux quantum. Second, minima are observed at $H = H_S + H_1$ (almost as deep as the absolute minimum for low currents), with $H_1 = \phi_0 / a^2 \approx 50$ Oe. These commensurability effects imply that for those fields the flux-lattice matches the square geometry of the array of "magnetic vortices" [5]. Third, $R(H)$ is strongly asymmetric: while commensurability effects are clear at $H = H_S + H_1$, they are barely observable at $H = H_S - H_1$. Moreover, the background resistance is larger for $H < H_S$ than for $H > H_S$.

Fig 3 (b) shows the magnetoresistance after an *out-of-plane* demagnetizing cycle (a series of minor loops of decreasing amplitude, from $H = 20$ kOe to $H = 0$). After this, a situation like the one in Fig. 1 (b) is expected. In this case, $R(H)$ curves are symmetric around $H = 0$ and no commensurability effects are observed.

In order to understand the behavior described above we need to consider the flux quanta induced by the external applied field, as well as those induced by the stray magnetic field from the dipoles ("magnetic vortex" cores). The total magnetic flux trough the film induced by a dipole is nearly zero [33]: as shown in the cartoon inside Fig. 2 (b), if a dipole points up the magnetic flux underneath the dipole is positive (field points up) whereas the same field lines create a negative magnetic flux around it. Under certain conditions and if the dipole is sufficiently strong, positive flux quanta $+\phi_0$ (either single quanta or a "giant" multiquanta) will be created and confined just underneath the dipole, and the same number of negative flux quanta $-\phi_0$ will appear arranged around it [20,35-37]. We will discuss later the actual situation in the studied samples. But now, let us assume that the magnetic stray field from the array of dipoles induces a certain number of flux quanta between them [38]. Fig. 4 shows a series of snapshots with the distribution of flux quanta between the dipoles as a function of the external applied field.

Fig. 4 (a)-(d) correspond to the case in which the array of dipoles is magnetized (i.e. all of them point "up") after the application and withdrawal of a large positive field. If the external field $H = 0$, the magnetic field lines from the dipoles "join" into negative flux quanta $-\phi_0$ between the dots. This is depicted in Fig. 4 (a) for the particular case in which there is 1/2 dipole-induced negative flux quantum $-\phi_0$ per unit cell of the array (the number of dipole-induced flux quanta between then dots may be different for different arrays, as we discuss later). Application of a positive (parallel to the dipoles) external field $H$ induces positive flux quanta $\phi_0$. These positive flux quanta annihilate [20] dipole-induced interstitial negative flux quanta $-\phi_0$. The absolute minimum resistance is observed when all of them are annihilated (Fig. 4 (b)) at $H_S$. This way, the annihilation of dipole-induced flux quanta leads to field-induced superconductivity. Further increase of the external field (above $H_S$) induces excess positive flux quanta $\phi_0$, and initially leads to an increase of the resistance until a second minimum develops at $H = H_S + H_1$. This corresponds to the well-known matching configuration [5-7,10] between the flux-lattice and the square array of dipoles (Fig. 4 (c), which leads to collective flux pinning. Conversely, application of a negative external field induces negative flux quanta $-\phi_0$. Because of their repulsive magnetostatic interaction with the dipoles, these are not pinned underneath them but stabilized in interstitial positions of the array, where they add to the dipole-induced flux quanta. For $H = H_S - H_1$ a shallow minimum is observed in the resistance. An ordered arrangement of flux quanta is expected at this field strength (Fig. 4 (d)), in which all the flux quanta are "caged" in interstitial positions by the surrounding dipoles. However, this sort of collective pinning is less effective than the one observed at $H = H_S + H_1$, for which flux quanta are sitting directly underneath the dipoles [6,7]. This asymmetric flux pinning gives rise to the asymmetry in Fig. 3 (a).

Fig. 4 (e)-(f) correspond to the demagnetized array. Although balanced, the distribution of polarities is probably disordered, as found in arrays with similarly large distances between vortex-cores [22,23]. When the external field $H = 0$ (Fig. 4 (e)), a few positive/negative flux quanta might be induced in areas of the array where there is a cluster of dipoles oriented in the same direction. At $H = H_1$ a situation like in Fig. 4 (f) is expected. Positive flux quanta are attracted to (repelled from) dipoles pointing "up" (down). As a result, only a fraction of the flux quanta are actually pinned by the dipoles, while for others interstitial positions are more favorable. This leads to a disordered flux-lattice as well as an adverse increase in the lattice elastic energy. Because of this, collective pinning does not develop and commensurability effects are not observed (Fig. 3 (b)). [39]

Finally, we discuss below the penetration of the field from the dipoles through the superconducting film, and estimate the size of the "magnetic vortex" cores. Using magnetostatics [40], we calculated the *out-of-plane* component of the magnetic field $H_\perp$ induced at the Al film plane by a "magnetic vortex" core. We assumed $M_\perp[r] = M_S(\Theta[s-r] + \Theta[r-s](r-R)/(s-R))$ to mimic the experimental magnetization profile in a Co vortex core [30], with $r$ the distance from the center of the core, $s = 2$ nm [30] the radius of the core section with maximum magnetization, $R$ the total core radius, $\Theta[x]$ the Heaviside step function, and $M_S = 1.43$ kOe the saturation magnetization. $M_\perp[r]$ for $R = 40$ nm is shown in the left inset of Fig. 5, and the induced $H_\perp(r)$ in its main panel. The flux of this field through the core area, $\phi$, is plotted in Fig. 5 (right inset) as a function of the core radius $R$, in units of $\phi_0$. Because of the partial screening provided by the Meissner currents (which depends on temperature and injected currents), the net "positive" flux *through the superconducting film* underneath a vortex core is $\phi_S^+ < \phi$ [20,35-37]. In an ordered array of vortex cores, in the absence of external fields, the net "positive" flux $\phi_S^+$ through the superconducting film *underneath* each of the cores equals the "negative" flux per unit cell $\phi_S^-$ through the area *between* the cores. [41] Thus, from the shift $H_S$ observed in $R(H)$, we calculated $\phi_S^+ = \phi_S^- = H_S a^2$ and obtained values $0.5\phi_0 \leq \phi_S^+ \leq \phi_0$.

From this and using $\phi(R)$ (right inset in Fig. 5) we estimate that the core radius $R > 40-60$ nm, in good agreement with experimental values for $R \sim 80$ nm for Co "magnetic vortices" [30].

As described above, we observed that the net field flux through the superconducting film *underneath* a dipole is $\phi_S^+ \leq \phi_0$. This is in contrast with previous findings for arrays of larger uniformly magnetized dots [32,34]. For these, $\phi_S^+$ jumps directly from $\phi_S^+ = 0$ to $\phi_S^+ > \phi_0$ and then increases in quantized steps $\phi_0$ as the dipole strength is continuously increased. I.e., in those experiments the shift $H_S$ of the superconducting/normal phase boundary (or shift of the $R(H)$ curves) is a multiple of the matching field $H_S = n\phi_0/a^2$, with $n$ and integer [32,34]. The different behavior observed in the present experiments may be caused by the different characteristics of the magnetic field profile from the vortex cores. On the one hand, the "positive" magnetic field underneath a vortex core is highly focused (see Fig. 5): it concentrates over a length scale $R$ smaller (much smaller) than the coherence length $\xi(0.84T_c) \sim 100$ nm (penetration length $\lambda(0.84T_c) \sim 550$ nm). On the other hand, due to the non uniform magnetization within the vortex core [30] (see left inset in Fig 5), the induced field is maximum underneath its center and decreases when approaching its peripheral (see Fig. 5). Contrary to this, in the case of larger uniformly magnetized dots, the magnetic field is nearly uniform in most of the area underneath the dot and peaks near its edges [36]. Further theoretical work is needed to check whether those differences in the field profiles result in significantly different distributions of screening currents and dipole-induced flux quanta over the array, and if these allow to explain the non-integer shift $H_S$ of the $R(H)$ curves ($\phi_S^+ < \phi_0$) observed in the present experiments.

**IV. CONCLUSION**

We have realized a superconducting/ferromagnetic hybrid system where the collective pinning of flux quanta can be switched *on* and *off* by manipulating the magnetic order of the ferromagnetic subsystem. This consists of an array of nanodots in the "magnetic vortex" state, in which *crucially* the nanodots have a permanent dipolar moment whose orientation can be manipulated via the magnetic history. In addition, we have observed asymmetric pinning and field-induced superconductivity effects. The latter originates from the annihilation of stray-field-induced flux quanta.

**ACKNOWLEDGEMENTS**

Work supported by NSF and AFOSR.

*Note added.* – At the time of resubmission of our manuscript we learned about the recent related publication by Hoffmann *et al.* [42]. In this work, the interaction of "magnetic vortices" and flux quanta is also studied. Hoffmann *et al.* suggest that the local suppression of superconductivity caused by the stray fields under the "magnetic vortices" is the governing flux pinning mechanism in their experiments. Contrary to this, the effects observed in our system indicate that magnetostatic interactions between "magnetic vortices" and flux quanta play a major role in flux pinning

FIGURE CAPTIONS

**Figure 1:** (Color online) (a) Ordered and (b) disordered array of magnetic dipoles on a superconducting film.

**Figure 2:** (Color online) (a) Lorentz microscopy image of a demagnetized array of Co dots with $a = 800$ nm and $\varnothing = 400$ nm, at room temperature. The vortex cores in the center of the dots appear as black (white) spots for clockwise (counter-clockwise) chirality. (b) Hysteresis loop at T=10 K of an array of Co dots with $a = 1000$ nm and $\varnothing = 450$ nm. Upper-left inset: sketch of a "magnetic vortex" and its stray magnetic field. Lower-left inset: cartoon of the magnetic reversal mechanism.

**Figure 3:** (Color online) Normalized magnetoresistance ($R_N$ normal-state resistance) at $T = 0.84 T_C$ with the field $H$ applied out-of-plane for a sample with Al thickness 10 nm, and array with $a = 600$ nm and $\varnothing = 490$ nm (a) after application and removal of a 20 kOe *out-of-plane* field and (b) after a demagnetizing cycle. Different line colors (types) for different injected currents (see legend in µA).

**Figure 4:** (Color online) Snapshot of the distribution of flux quanta over the array of magnetic dipoles, as a function of the external applied field $H$ for (a)-(d) a magnetized array and (e)-(f) a demagnetized array. Magnetic vortex cores pointing up (down) are depicted by light crossed (dark dotted) circles. Positive (negative) flux quanta are depicted by light (dark) areas encircled by counterclockwise (clockwise) circulating arrows. These arrows mimic the sense of circulation of supercurrents.

**Figure 5:** (Color online) Out-of-plane component of the magnetic field $H_\perp$ induced by a "magnetic vortex" core of radius $R = 40$ nm, as a function of the distance to its center $r$. The horizontal line points out the experimental $H_{c2}(0.84 T_c) \approx 300$ Oe for the sample in Fig. 3, determined from magnetotransport measurements. Inset (a): out-of-plane magnetization $M_\perp$

profile within a "magnetic vortex" core (see text). Inset (b): Flux of $H_\perp$ through the "magnetic vortex" core area as a function of the core radius $R$, in units of the flux quantum $\phi_0$.

*local* matching with ordered "clusters" of dipoles disperse across the demagnetized array, similarly to *local* matching observed with some quasiperiodic arrays of pinning centers [12].

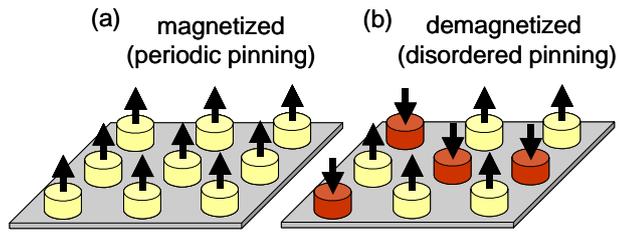

Figure 1

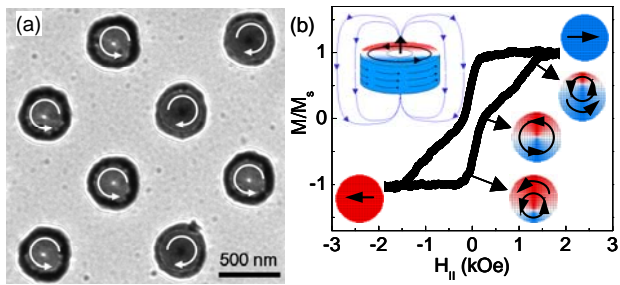

Figure 2

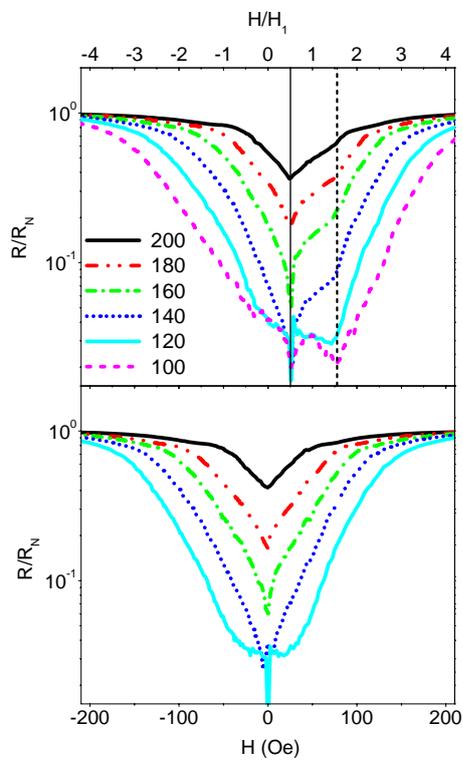

Figure 3

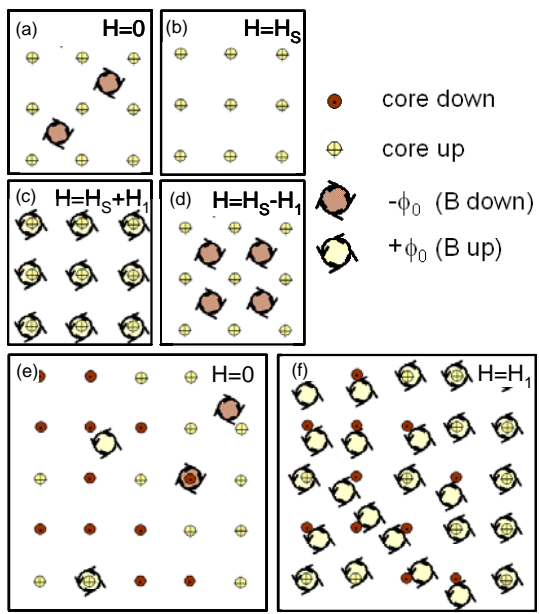

Figure 4

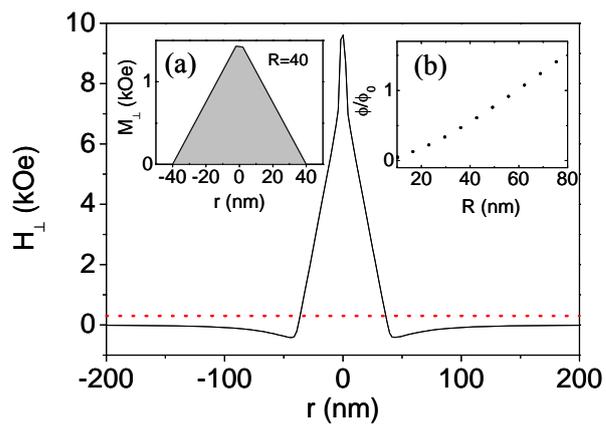

Figure 5